\documentclass[apj]{emulateapj}
\slugcomment{{\sc Accepted for publication in The Astrophysical Journal:} December 4, 2010}

\newcommand{\rpfifty}{$r_{p} \leq 50~h^{-1}$ kpc }
\newcommand{\rpthirty}{$r_{p} \leq 30~h^{-1}$ kpc }

\usepackage{amsmath}
\usepackage{graphicx}
\usepackage{epsfig}
\usepackage{multirow}
\usepackage{wrapfig}
\shorttitle{PRIMUS: ENHANCED SSFR IN CLOSE PAIRS}
\shortauthors{WONG ET AL.}

\begin{document}
\title{PRIMUS: ENHANCED SPECIFIC STAR FORMATION RATES IN CLOSE GALAXY PAIRS}
\author{
Kenneth C. Wong\altaffilmark{1},
Michael R. Blanton\altaffilmark{2},
Scott M. Burles\altaffilmark{3},
Alison L. Coil\altaffilmark{4,8},
Richard J. Cool\altaffilmark{5,9},
Daniel J. Eisenstein\altaffilmark{1,6},
John Moustakas\altaffilmark{4},
Guangtun Zhu\altaffilmark{2},
and St\'{e}phane Arnouts\altaffilmark{7}}
\altaffiltext{1}{Steward Observatory, University of Arizona, 933 North Cherry Avenue, Tucson, AZ 85721}
\altaffiltext{2}{Center for Cosmology and Particle Physics, Department of Physics, New York University, 4 Washington Place, New York, NY 10003}
\altaffiltext{3}{D.E. Shaw \& Co., L.P., 20400 Stevens Creek Blvd., Suite 850, Cupertino, CA 95014}
\altaffiltext{4}{Department of Physics, University of California, 9500 Gilman Dr., La Jolla, San Diego, CA 92093}
\altaffiltext{5}{Department of Astrophysical Sciences, Princeton University, Peyton Hall, Princeton, NJ 08544}
\altaffiltext{6}{Harvard-Smithsonian Center for Astrophysics, 60 Garden Street, Cambridge, MA 02138}
\altaffiltext{7}{Canada-France-Hawaii Telescope Corporation, 65-1238 Mamalahoa Hwy, Kamuela, HI 96743}
\altaffiltext{8}{Alfred P. Sloan Foundation Fellow}
\altaffiltext{9}{Hubble Fellow and Carnegie-Princeton Fellow}

\begin{abstract}
Tidal interactions between galaxies can trigger star formation, which contributes to the global star formation rate density of the universe and could be a factor in the transformation of blue, star-forming galaxies to red, quiescent galaxies over cosmic time.  We investigate tidally-triggered star formation in isolated close galaxy pairs drawn from the Prism Multi-Object Survey (PRIMUS), a low-dispersion prism redshift survey that has measured $\sim$120,000 robust galaxy redshifts over 9.1 deg$^{2}$ out to $z \sim 1$.  We select a sample of galaxies in isolated galaxy pairs at redshifts $0.25 \leq z \leq 0.75$, with no other objects within a projected separation of $300~h^{-1}$ kpc and $\Delta z/(1+z)$ = 0.01, and compare them to a control sample of isolated galaxies to test for systematic differences in their rest-frame $FUV-r$ and $NUV-r$ colors as a proxy for relative specific star-formation rates.  We find that galaxies in \rpfifty pairs have bluer dust-corrected $UV-r$ colors on average than the control galaxies by $-0.134 \pm 0.045$ magnitudes in $FUV-r$ and $-0.075 \pm 0.038$ magnitudes in $NUV-r$, corresponding to a $\sim 15-20$\% increase in SSFR.  This indicates an enhancement in specific star formation rate due to tidal interactions.  We also find that this relative enhancement is greater for a subset of \rpthirty pair galaxies, for which the average colors offsets are $-0.193 \pm 0.065$ magnitudes in $FUV-r$ and $-0.159 \pm 0.048$ magnitudes in $NUV-r$, corresponding to a $\sim 25-30$\% increase in SSFR.  We test for evolution in the enhancement of tidally-triggered star formation with redshift across our sample redshift range and find marginal evidence for a decrease in SSFR enhancement from $0.25 \leq z \leq 0.5$ to $0.5 \leq z \leq 0.75$.  This indicates that a change in enhanced star formation triggered by tidal interactions in low density environments is not a contributor to the decline in the global star formation rate density across this redshift range.
\end{abstract}


\section{INTRODUCTION} \label{sec:intro}
Interactions between galaxies are an important process in the evolution of galaxies with cosmic time, as they can affect various galaxy properties such as star formation rate, morphology, and gas fraction.  Simulations have shown that major interactions and mergers between galaxies can produce disturbed morphologies and trigger starbursts \citep[e.g.][]{toomre1972,barnes1991,barnes1992,mihos1992,kauffmann1993,mihos1994,mihos1996,springel2000,tissera2002,bundy2005,cox2006,dimatteo2007,dimatteo2008,lotz2008}.  Interacting and merging systems can also trigger intense infrared emission in gas-rich galaxies, resulting in the formation of luminous and ultra-luminous infrared galaxies \citep[LIRGs and ULIRGs;][and references therein]{sanders1996}.  The increase in the incidence of LIRGs and ULIRGs at intermediate redshifts \citep{lefloch2005,rujopakarn2010} suggests that the decline in the global star formation rate since redshift $z \sim 1$ \citep{lilly1996,madau1996,chary2001,perezgonzalez2005} may be caused by a decrease in the amount of star formation triggered by interactions over time.

Past studies have shown that star formation triggered by major mergers is not a significant fraction of the overall star formation rate at intermediate redshifts, however.  Using data from the Galaxy Evolution from Morphology and SEDs survey \citep[GEMS;][]{rix2004} and Classifying Objects by Medium-Band Observations in 17 Filters \citep[COMBO-17;][]{wolf2001,wolf2004} survey, \citet{wolf2005} found that morphologically-identified merging galaxies contribute roughly 20\% of the ultraviolet luminosity density (and thus, star formation rate density) at $z \sim 0.7$.  \citet{robaina2009} used data from GEMS and COMBO-17, along with data from the Space Telescope A901/2 Galaxy Evolution Survey \citep[STAGES;][]{gray2009}, to show that $\lesssim 10$\% of star formation at $0.4 \leq  z \leq 0.8$ is triggered by these major interactions.  The conclusions drawn from these observational studies are in general agreement with the results of simulations performed by \citet{hopkins2010}, who find that only $\sim5-10$\% of the star formation rate density out to $z \sim 6$ is the result of merger-induced starbursts.

While star formation triggered by major interactions and mergers may not be a significant contributor to the evolution of the global star formation density, it is also important to examine the properties of the likely progenitor population $-$ close galaxy pairs \citep[e.g.][]{patton2000,depropris2007} $-$ to investigate whether more frequent tidal interactions in close pairs could play a role.  Searches for close galaxy pairs have been used by many independent studies to derive the evolution of galaxy merger rates out to intermediate redshifts \citep[e.g.][]{burkey1994,carlberg1994,patton1997,patton2002,lefevre2000,lin2004,lin2008,bell2006,kartaltepe2007,deravel2009}.  Star formation triggered by these tidal interactions may also consume much of the galaxies' cold gas, which could be a key factor in the transformation of blue, star-forming galaxies to red, quiescent galaxies.  We expect it would be a bigger factor in dense environments where these interactions are more common, so tidally-triggered star formation may also be an important contributor to the redshift evolution of galaxies in different environments.

The first indication that tidal interactions could affect galaxy properties was found by \citet{larson1978}, who observed that interacting systems identified by peculiar morphologies showed much greater scatter in their optical color distributions in comparison to morphologically-normal galaxies.  This was interpreted to be the result of a recent burst of star formation triggered by the tidal interaction.  Subsequent studies \citep[e.g.][]{condon1982,keel1985,kennicutt1987} found similar results indicating that rapid bursts of star formation were associated with tidal interactions between galaxies.  Most studies of tidally-triggered star formation in close galaxy pairs have been performed at low redshift ($z \lesssim 0.1$) using H$\alpha$ emission as the main diagnostic of star formation rate.  These studies have indicated that there is an overall enhancement in the star formation rate of close pairs of galaxies relative to a similar population of isolated galaxies, and that the strength of the enhancement is anticorrelated with the pair separation.  One of these early studies by \citet{kennicutt1987} found an enhancement in both H$\alpha$ and far-infrared emission in interacting galaxies compared to isolated galaxies, accounting for $\sim 6$\% of the total massive-star formation in luminous star-forming galaxies.  However, they found that the degree of enhancement had a large variation and that they were strongly biased toward unusually bright and active systems.  Thus, they were unable to draw conclusions about more typical interacting galaxy pairs.

In order to isolate the effects of star formation triggered by tidal interactions, it is important to take into account the local environments of the galaxies in question.  Dense environments, such as galaxy groups and clusters, are the most likely locations of tidal galaxy-galaxy interactions.  However, these dense environments have a higher fraction of red elliptical galaxies with little ongoing star formation \citep{dressler1980,postman1984,hermit1996,guzzo1997,giuricin2001,gomez2003,kauffmann2004,blanton2005}.  It has been shown that tidally-triggered star formation in galaxy pairs is more pronounced at low densities, likely due to the fact that galaxies in dense environments have had their gas content exhausted by previous interactions \citep{solalonso2006}.  \citet{barton2007} used simulations to find that a significant fraction of close pairs tended to lie in host dark matter halos that contained more galaxies than just the two in the pair.  They established that in order to isolate triggered star formation, a sample of isolated pairs must be constructed and compared to a sample of isolated galaxies in the same sparse environments.  \citet{perez2009a} investigated the effects of various sources of bias that could influence the results derived from a comparison of galaxy pairs to a control sample.  Using semi-analytical models, they showed that the local density in the environment of close pairs and the control galaxies is one of the most significant sources of bias and must be accounted for when selecting an appropriate control sample.

For observational studies, selecting a large sample of galaxy pairs in low-density regions is difficult due to the low frequency of close pairs that lie in these environments.  Only large-scale redshift surveys such as the Sloan Digital Sky Survey \citep[SDSS;][]{york2000} and 2dF Galaxy Redshift Survey \citep[2dFGRS;][]{colless2001} can provide datasets with sufficient volume for statistical studies of galaxy pairs in a variety of environments.  The spectroscopic data from these surveys can be used to create a sample of true isolated galaxy pairs by filtering out interloping apparent pairs that are close in projected separation but far apart in redshift, as well as deprojecting nearby galaxies to get a better handle on their local environments.

More recent galaxy pair studies \citep{nikolic2004,woods2007,ellison2008,li2008} have taken advantage of these larger samples from the SDSS and found an anticorrelation between specific star formation rate (SSFR) and projected separation, $r_{p}$, between the pairs within $\sim 30~h^{-1} - 100~h^{-1}$ kpc.  These works do not explicitly account for the local environments of the pair and control galaxies, allowing them to define large pair samples containing several thousand galaxies.  \citet{li2008} did test the relative enhancements in isolated and non-isolated pairs and did not find a significant difference.  However, they characterize the environment only in a very small region (projected within $100~h^{-1}$ kpc) around their galaxies, which the results of \citet{barton2007} suggests is not large enough to accurately quantify the local density.  \citet{nikolic2004} and \citet{woods2007} both find a stronger enhancement in tidally-triggered star formation for blue star-forming galaxies.  Interestingly, \citet{nikolic2004} and \citet{li2008} find little dependence of the enhancement on the relative mass or luminosity of the pair galaxies to their companions, whereas \citet{woods2007} and \citet{ellison2008} find that the relative magnitude of the pair is important in the effectiveness of the tidally-triggered interactions.

A similar dependence of star formation rate on projected separation in close pairs has been found by \citet{barton2000} and \citet{woods2006} in the CfA2 redshift survey \citep{geller1989}, and by \citet{lambas2003} in the 2dFGRS.  While these studies have made use of large-volume surveys at low redshifts to investigate a large sample of close pairs, comparatively few studies of tidal interactions have been performed at intermediate redshifts, where the effects may be more prominent due to the higher star formation rate density and gas fraction in the universe at those epochs.  These intermediate-redshift studies could be valuable in determining whether a decline in star formation triggered by tidal interactions contributes to the decline in the global star formation rate.

Only recently have intermediate-redshift surveys with sufficient volume been utilized to examine close pairs in this regime.  \citet{lin2007} used data from the DEEP2 survey \citep{davis2003} along with infrared fluxes from Spitzer and found an enhanced SSFR in \rpfifty close pairs and morphologically-identified merger systems, but were unable to draw conclusions about the redshift evolution of the signal between redshifts 0.1 and 1.1.  They also found that the anticorrelation between star formation rate and pair separation seen in low-redshift pair studies was seen at higher redshift.  \citet{deravel2009} found an enhancement in [O II] luminosity in close pairs out to $z \sim 1$ in the VIMOS VLT Deep Survey \citep[VVDS;][]{lefevre2005}, although they did not control for the local environment of the pairs and focused primarily on the evolution of the merger rate.  \citet{woods2010} studied the enhancement of SSFR in the Smithsonian Hectospec Lensing Survey \citep[SHELS;][]{geller2005} at redshifts $0.08 \leq z \leq 0.38$ and found similar trends to the previous low-redshift studies, but they also did not look for redshift evolution in their sample.

In this paper, we investigate the relative SSFR of galaxies in isolated close pairs compared to a control sample of isolated galaxies in the Prism Multi-Object Survey \citep[PRIMUS\footnote{http://cass.ucsd.edu/$\sim$acoil/primus/};][Cool et al. in preparation]{coil2010}, a low-dispersion prism intermediate-redshift survey that provides the largest sample of faint galaxy redshifts yet to $z \sim 1$.  We use existing UV and optical photometry to infer rest-frame $UV-r$ colors as a proxy for SSFR.  One advantage of studying galaxies at intermediate redshifts is that the ultraviolet emission, which measures the young intermediate-to-high-mass ($\gtrsim 5~M_{\odot}$) stellar population of galaxies, is shifted redward, close to or into observed-frame optical wavelengths.  This makes it an ideal tracer of star formation over a range of redshifts \citep[][and references therein]{kennicutt1998}, although the UV is more susceptible to attenuation by dust.  We have deblended UV data from the Galaxy Evolution Explorer \citep[$GALEX$;][]{martin2005} that we use to determine rest-frame fluxes in both the far-UV ($FUV$; $\lambda_{eff} \sim 1530$ \AA) and near-UV ($NUV$; $\lambda_{eff} \sim 2270$ \AA) bands.

Our goal is to study the relative effects of tidal interactions on SSFR at intermediate redshifts ($z \sim 0.5$), where relatively few studies of this nature have been performed in comparison to low redshift studies.  We investigate galaxies in pairs with a projected separation of \rpfifty, which has been used in previous studies \citep[e.g.][]{barton2000,woods2010} as a typical distance within which to define interacting systems.  We also select a subsample of galaxies in pairs with a projected separation of \rpthirty to look for an increased enhancement in SSFR with decreasing separation, as has been indicated by past studies.  We focus on the SSFR rather than the absolute star formation rate so that we are not biased toward intrinsically more massive and luminous galaxies with increasing redshift.  We examine how those effects might evolve over a range of redshifts ($0.25 \leq z \leq 0.75$) to investigate whether the overall downward trend of star formation rate in field galaxies is reflected in isolated pair galaxies undergoing tidal interactions.  A large redshift survey like PRIMUS ($\sim$120,000 robust galaxy redshifts over 9.1 deg$^{2}$ out to $z \sim 1$) is needed for such a study in order to define a clean sample of isolated pair galaxies that is large enough to compare statistically to an unbiased control sample.

This paper is organized as follows.  In \S~\ref{sec:data}, we describe the PRIMUS dataset and the cuts we apply to create a clean sample from which to select pair and isolated galaxies.  In \S~\ref{sec:color}, we describe our methodology for determining the $FUV$ and $NUV$-band fluxes for our sample.  We describe our method for selecting isolated pair galaxies and a corresponding control sample in \S~\ref{sec:sample}.  We present our results in \S~\ref{sec:results} and discuss them in \S~\ref{sec:discussion}.  We summarize our main conclusions in \S~\ref{sec:conclusions}.  Throughout this paper, we assume a $\Lambda$CDM cosmology with $\Omega_{m} = 0.3$, $\Omega_{\Lambda} = 0.7$, and $h = 0.71$.  All magnitudes used in this paper are on the AB system.  Rest-frame colors are denoted by a preceding superscript zero, e.g. $^{0}(u-r)$.

\section{DATA} \label{sec:data}
We draw our sample of galaxies from the Prism Multi-Object Survey \citep[PRIMUS;][Cool et al. in preparation]{coil2010}, a low-resolution spectroscopic redshift survey covering 9.1 deg$^{2}$ of the sky with existing multiwavelength data from the infrared to the X-ray, including UV data from $GALEX$.  The PRIMUS catalog contains robust redshifts for $\sim$120,000 objects to a flux limit of $i \sim 23$.  The redshifts from PRIMUS have been demonstrated to be accurate to within $\sigma_{z}/(1+z) \lesssim 0.005$ and contain $\lesssim 3$\% catastrophic outliers ($\Delta z/(1+z) \geq 0.03$).  The advantage of using the PRIMUS dataset is that it surveys a volume comparable to the 2dFGRS, but at redshifts out to $z \sim 1$.  By sacrificing spectral resolution for throughput, PRIMUS is able to observe $\sim 2,500$ objects simultaneously and probe a much larger volume than high-resolution spectroscopic surveys at comparable redshifts such as DEEP2 and VVDS.

PRIMUS targets multiple independent fields covering a total area of 9.1 deg$^{2}$.  The targeted fields have multiple bands of optical photometry that were used in the redshift determination.  The full details of the optical photometry, zeropoints, and relevant references are given in \citet{coil2010}.

\begin{figure}
\centering
\plotone{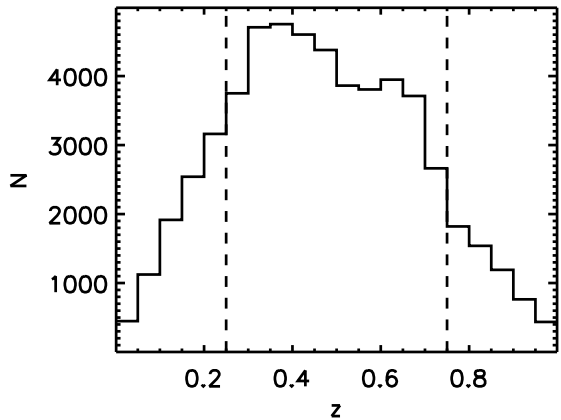}
\caption{Redshift distribution of the parent sample.  The dashed lines indicate the redshift range ($0.25 \leq z \leq 0.75$) from which we select the pair sample.  \label{fig:parent_z_hist}}
\end{figure}

To ensure a high-quality sample from which to select pair galaxies and control samples, we apply a series of selection criteria to the PRIMUS catalog.  First, we only include objects that have a redshift confidence of $Q = 4$, which creates the most pure sample with the lowest number of redshift outliers, and we also select objects that do not have any extraction flags indicating poor data quality or an uncertain redshift determination (see Cool et al. in preparation for details).  We also require objects to have been identified as a galaxy (as opposed to a star or broad-line AGN) by the templates used in the redshift fitting.  In addition, we only include objects that have an apparent $i$-band magnitude (inferred by K-correcting from the nearest available photometric band to the observed-frame $i$-band) brighter than 22.5.  The catastrophic outlier rate for galaxies brighter than this limit is $\lesssim5$\% but has not been well-tested at fainter magnitudes, so we do not include those objects in our analysis.  Finally, we only include objects in regions where we have deblended $GALEX$ fluxes (see \S~\ref{subsec:uv_flux}).  Hereafter, we refer to this pruned catalog as the ``parent sample''.  The parent sample contains 55,944 galaxies, and its redshift distribution is shown in Figure~\ref{fig:parent_z_hist}.

All rest-frame quantities are computed using the kcorrect software package \citep[v4.1.4;][]{blanton2007}.  To summarize, the software fits the sum of a set of basis templates at the PRIMUS redshift to the broadband optical photometry available in each field.  The basis templates are based on stellar population synthesis models and are constrained to produce a non-negative best-fit template.  It is this best-fit template that we use to infer rest-frame absolute magnitudes by K-correcting from the nearest observed photometric band in that object's particular field.

\section{DETERMINATION OF UV-r COLOR} \label{sec:color}
We quantify the SSFR of galaxies by calculating their rest-frame $FUV-r$ and $NUV-r$ colors.  \citet{salim2005} has shown that $NUV-r$ is tightly correlated with the ratio of current (within the past $\sim$100 Myr) to past-averaged star formation, which is related to SSFR \citep{kennicutt1994}.  The UV flux is a proxy for star formation rate since it traces young, massive stars, and the $r$-band flux is a proxy for stellar mass since it will trace the older stellar populations of galaxies and is relatively insensitive to extinction from the interstellar medium of the host galaxy.

\subsection{UV Flux Determination} \label{subsec:uv_flux}
$GALEX$ has a typical PSF full-width at half-maximum of 4.2\arcsec~in the $FUV$-band and 5.3\arcsec~in the $NUV$-band \citep{morrissey2007}.  Although deep $GALEX$ photometry exists for all of the PRIMUS fields, these PSF widths correspond to a projected distance ranging from $\sim15~h^{-1} - 25~h^{-1}$ kpc at the redshift considered in our analysis, which is problematic for resolving close pairs with smaller projected separations.  Therefore, a team member (S. Arnouts) has created a catalog of object fluxes from $GALEX$ that have been determined using a Bayesian deblending technique \citep{guillaume2006}.  The technique uses an expectation-maximization (EM) algorithm to determine flux solutions for blended objects using optical priors from the bluest available photometric band in each field (typically $u$-band or $B$-band) and has been used in past studies with deep $GALEX$ imaging \citep[e.g.][]{zamojski2007,ilbert2009,salim2009,hammer2010}.  The EM algorithm is only applied to objects fainter than $NUV = 21.75$.  At brighter magnitudes, blending and confusion is not a major issue, so the GALEX pipeline photometry is used.  GALEX coverage exists for 7.47 of the 9.1 deg$^{2}$ observed by PRIMUS.  For this analysis, we only include objects that are in the GALEX catalog.

\subsection{UV Extinction Corrections} \label{subsec:ext_corr}
Actively star-forming galaxies are expected to have their UV flux attenuated due to dust \citep{sanders1996,kennicutt1998}.  As a result, UV fluxes will be underestimated unless dust extinction is accounted for.  We use the relations determined by \citet{cortese2008} to convert the galaxies' total infrared-to-UV (TIR/UV) ratio and $^{0}(UV-r)$ color into extinction corrections for both the $FUV$-band and $NUV$-band, $A_{FUV}$ and $A_{NUV}$, respectively.  These relations use the $^{0}(UV-r)$ colors to control for the age of the underlying stellar population.  If unaccounted for, the UV attenuation calculated from other methods \citep[e.g.][]{kong2004,buat2005,salim2007} will be overestimated in general.

We estimate the total infrared-to-FUV ratio (TIR/FUV) of a galaxy using its $^0(FUV-NUV)$ color \citep{cortese2006,cortese2008} for the $A_{FUV}$ calculation.  In calculating $A_{NUV}$, we convert this into a TIR/NUV ratio.  We note that \citet{cortese2008} claim that these extinction corrections are valid only for galaxies with $^{0}(FUV-r) \geq 1.6$ and $^{0}(NUV-r) \geq 1.5$, respectively.  For galaxies with bluer colors, we will underestimate the extinction correction needed.  However, only $\sim20$\% of the galaxies in both our pair and control samples fall in this range, so the extinction correction should be valid for the majority of our sample.

Given the number of assumptions made in determining this extinction correction, we investigate both the corrected and uncorrected $UV-r$ colors in our results (\S~\ref{sec:results}).

\section{SAMPLE SELECTION} \label{sec:sample}
Simulations have shown that the fraction of galaxies in close pairs exhibiting enhanced star formation relative to isolated galaxies can be biased toward lower values if the environment of the interacting galaxy is not taken into account \citep{barton2007,perez2009a}.  This effect results from the well-known morphology-density relation: a larger fraction of galaxies in overdense environments are red early-type ellipticals with little star formation compared to galaxies in underdense environments, where there is a larger fraction of blue late-type spirals \citep{dressler1980,postman1984,giuricin2001,gomez2003,kauffmann2004,blanton2005}.  The results of simulations by \citet{barton2007} show that galaxies in \rpfifty close pairs are likely to be in dark matter halos that contain more galaxies than the two comprising the pair, and in many cases ($\sim$40\%) contain 9 or more galaxies.  These dense environments will have large effects on the inferred SSFR of the galaxy pair.  In order to distinguish star formation triggered by tidal interactions with a single companion galaxy from environmental effects, one must focus on pairs in low density environments and compare the relative SSFR of those galaxies with that of a control sample of galaxies in a similar environment.

The simulations of \citet{barton2007} show that the effects of a tidal interaction between members of a galaxy pair can be separated from the effects of the surrounding environment by considering only pairs with no other objects within a projected separation of $700~h^{-1}$ kpc, and comparing them to a control sample consisting of isolated galaxies with no other objects within $300~h^{-1}$ kpc and at most one object between $300~h^{-1}$ kpc and $700~h^{-1}$ kpc away.  However, they find that the majority ($\sim$70\%) of galaxies in their simulations with a projected separation of at least $300~h^{-1}$ kpc to their nearest neighbors are in isolated halos, and this fraction increases with even larger projected distances to their nearest neighbor.  They also find that the masses of isolated close pair host halos are generally two to three times the masses of isolated single galaxy host halos, indicating that isolated galaxies are the progenitors of the isolated pairs.  Since the local environment of a progenitor should remain the same after an interaction with another isolated galaxy, one would expect that pairs with no other galaxies within $300~h^{-1}$ kpc should also mostly reside in host halos with no galaxies other than itself and its companion.

The \citet{barton2007} simulations were also designed to match a volume-limited sample of $M_{B} \leq -19$ galaxies selected from the 2dFGRS.  Our parent sample, on the other hand, is a flux-limited sample.  The $B$-band luminosity for galaxies at our detection limit ranges from $M_{B}-5\log{h} \approx -16$ at $z = 0.25$ to $M_{B}-5\log{h} \approx -18.5$ at $z = 0.75$.  Roughly one-third of our pair and isolated sample consists of galaxies less luminous than those used in the \citet{barton2007} analysis.  We conclude that their results, when applied to our sample, are overly conservative in the sense that had they considered less luminous galaxies, they likely would have found that they could have set their projected separation threshold to less than $700~h^{-1}$ kpc and still have found most of their pair and isolated galaxies in isolated halos.  Therefore, we must reduce our cut on the projected distance to the nearest neighbor since our sample includes fainter galaxies.  For these reasons, we apply a threshold of $ 300~h^{-1}$ kpc in radius to define our isolated pair and control samples.  We test for potential biases resulting from our use of a flux-limited sample in the discussion of our results (\S~\ref{sec:discussion}).

\subsection{Pair Sample} \label{subsec:pair_sample}
The sample of isolated pair galaxies (hereafter referred to as the ``pair sample'') is defined as objects in the parent sample with exactly one neighbor within $50~h^{-1}$ kpc in projected separation and within $\Delta z/(1+z) \leq 0.01$, with no other objects within a projected separation of $300~h^{-1}$ kpc and within $\Delta z/(1+z) \leq 0.01$ of the mean redshift of the pair.  We also consider a subset of these pair galaxies that have a projected separation less than $30~h^{-1}$ kpc.  The redshift-space cut corresponds to a recessional velocity difference of $\Delta V \leq 3000$ km s$^{-1}$.  This is a larger $\Delta V$ cut than past studies have used (typically $\Delta V \sim 500 - 2000$ km s$^{-1}$) to account for peculiar velocities, but is necessary due to the redshift uncertainty of PRIMUS.  This will have the effect of introducing more interlopers into our pair sample and will dilute any signal from triggered star formation, making our results a lower limit.

\begin{figure*}
\centering
\plotone{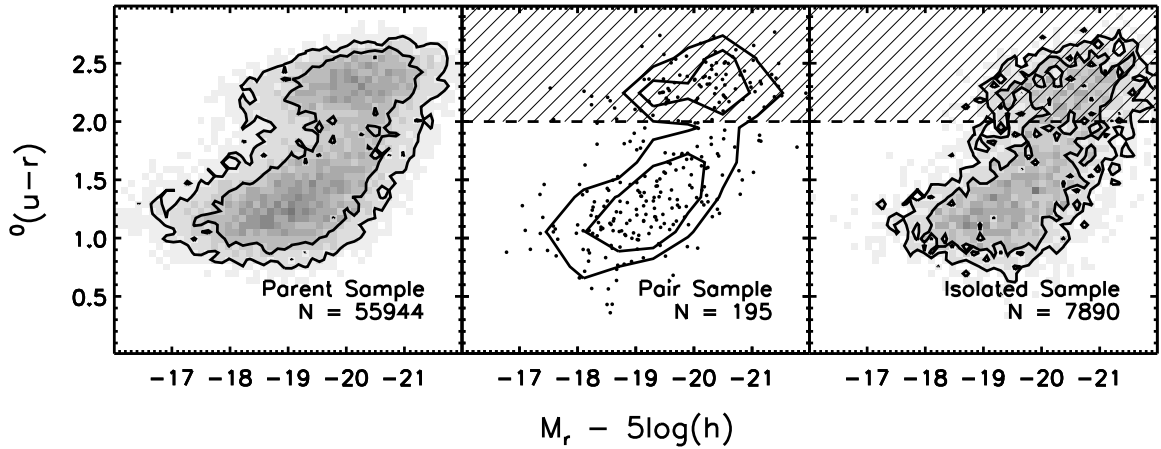} 
\caption{Color-magnitude diagram for the parent sample (left), isolated pair galaxies with \rpfifty (center), and isolated sample (right).  The 50\% and 75\% contours are shown for clarity.  The hatched region indicates the pair and isolated galaxies with $^{0}(u-r) > 2$, which we exclude from our analysis.  The number of objects in each sample, not including the hatched region, is given in the bottom right corner of the corresponding panel. \label{fig:cmd}}
\end{figure*}

We impose a lower limit of $r_{p} \geq 5~h^{-1}$ kpc on the projected separation of pairs in our sample since at smaller separations it is difficult to distinguish true merging systems from spurious pair detections (e.g. irregular galaxies that may be misidentified in the photometry as two separate objects).  \citet{patton2000} find that these very close pairs should only account for $\sim$5\% of true $r_{p} \leq 20~h^{-1}$ kpc pairs based on an extrapolation of the correlation function, so they should comprise an even smaller fraction of our pair sample.  We limit the pair sample to objects where at least 90\% of the area projected within $300~h^{-1}$ kpc of the object falls within the observed PRIMUS region.  This removes objects that lie too close to the boundaries of the survey (including CCD gaps) or to a region masked due to a bright star since we will be limited in our ability to constrain the local environment in those areas.  We only include objects that have no more than two potentially conflicting neighbors in the PRIMUS targeting catalog (defined as nearby objects in projection that overlap with the $30\arcsec\times8\arcsec$ extraction region of the galaxy).  This removes objects for which we might potentially underestimate the density of its local environment due to density-dependent sampling in PRIMUS (see \citet{coil2010} for details), although many of these objects may actually be at different redshifts than the pair galaxy of interest.  We also only consider galaxies in our full sample with $^{0}(u-r) \leq 2$ (Figure~\ref{fig:cmd}), which approximately selects galaxies in the blue cloud.  This is done to ensure that we are not comparing galaxies from completely different populations.  Based on past results, we expect the blue galaxies to show a more pronounced signal \citep{nikolic2004,solalonso2006,woods2007}.  This optical color selection potentially removes dusty star-forming galaxies on the red sequence, but only a small fraction ($\sim$15\%) of red sequence galaxies out to $z = 0.5$ at a flux limit similar to that of our sample is actively star-forming \citep{zhu2010}.  We do not place constraints on the luminosity difference between galaxies in a pair, so both major and minor interactions are included in our sample.

\begin{figure*}
\centering
\plotone{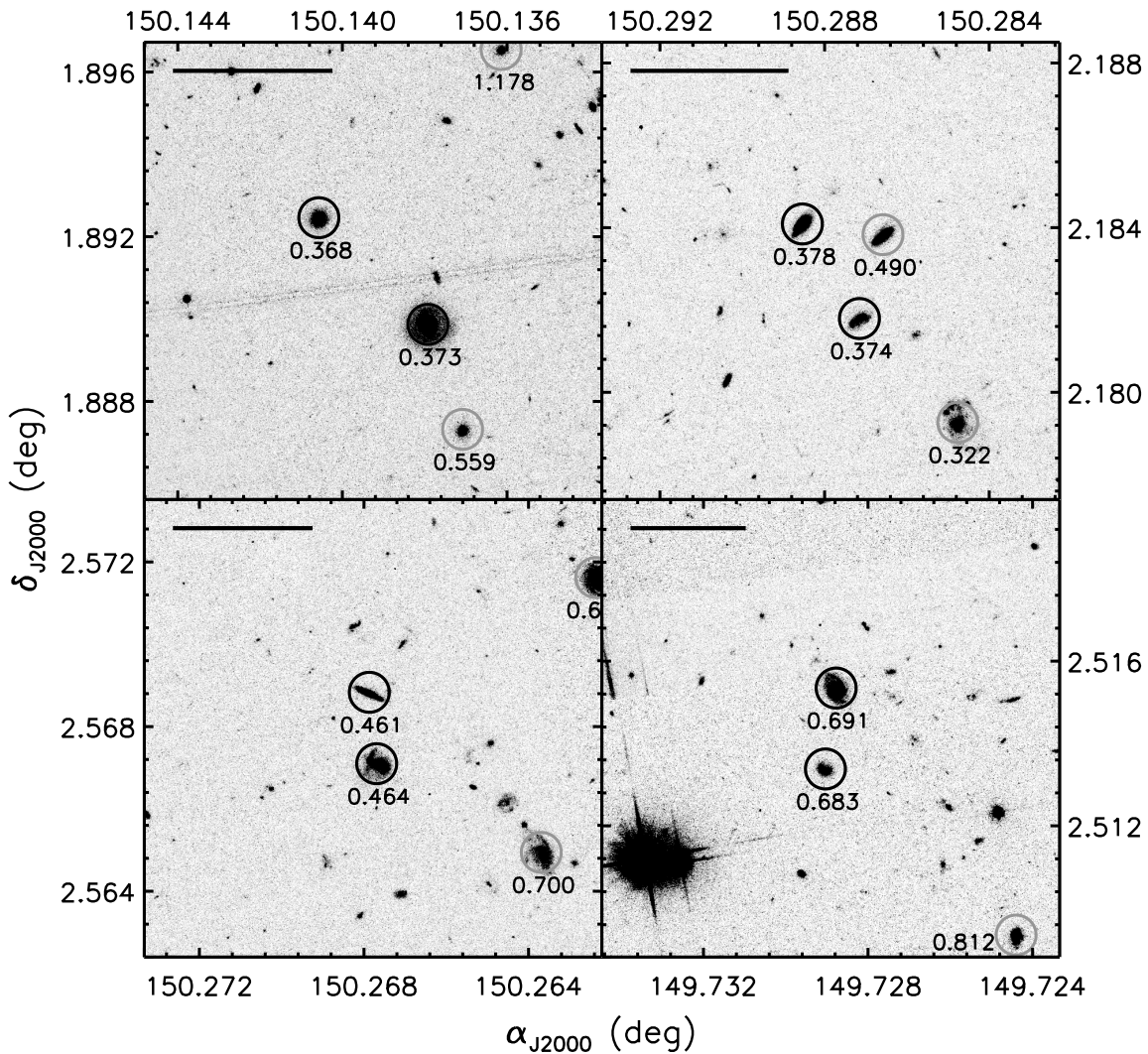}
\caption{HST imaging of a subset of close galaxy pairs  in the COSMOS field.  Circles have been drawn around the galaxies in the pair (black) and other objects in the parent sample (grey).  The redshift of each object is indicated below the circle.  The length of the bar in the upper left corner of each panel represents a projected distance of $50~h^{-1}$ kpc at the mean redshift of the depicted pair.  \label{fig:cosmos_pairs}}
\end{figure*}

We note that these selection criteria are applied to pair galaxies individually so that it is possible for one galaxy in a pair to be included in the pair sample, while its companion is not.  We show HST imaging for a subset of pairs in Figure~\ref{fig:cosmos_pairs} as an example of typical galaxy pairs in our sample.  These pairs are taken from the COSMOS field, where existing HST imaging is publicly available.

\begin{figure}
\centering
\plotone{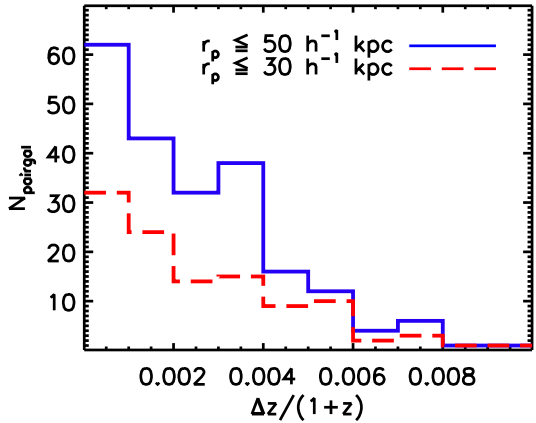}
\caption{Redshift-space pair separation distribution for \rpfifty pair galaxies (blue solid line) and \rpthirty pair galaxies (red dashed line).  Roughly 90\% of our pair galaxies are separated from their neighbor by $\Delta z / (1+z) \leq 0.005$ ($\Delta V \leq 1500$ km s$^{-1}$).   \label{fig:pair_dz_hist}}
\end{figure}

Due to the lack of volume at low redshift in PRIMUS and the drop in the observed number density at higher redshift (see Figure~\ref{fig:parent_z_hist}), we restrict the pair sample to galaxies in the range $0.25 \leq z \leq 0.75$.  After applying these cuts, we have a total of 195 \rpfifty pair galaxies.  Of these pair galaxies, 101 are in \rpthirty pairs.  In Figure~\ref{fig:pair_dz_hist}, we plot the distribution of $\Delta z / (1+z)$ for our pair sample.  We find that although we chose a relatively large cut at $\Delta z / (1+z) \leq 0.01$ ($\Delta V \leq 3000$ km s$^{-1}$) to allow for redshift uncertainties, roughly $90$\% of the galaxies in the pair sample have a redshift-space separation within $\Delta z / (1+z) \leq 0.005$ of their neighbor, corresponding to a recessional velocity difference within $\Delta V \leq 1500$ km s$^{-1}$.  The median recessional velocity difference for the pair sample is $\Delta V \sim 635$ km s$^{-1}$.

In order to test for any possible redshift dependence of a detected signal, we subdivide our pair sample into two bins in redshift: a lower-redshift ($0.25 \leq z \leq0.5$) and higher-redshift ($0.5 \leq z \leq0.75$) bin.  We select a separate control sample (see \S~\ref{subsec:control_sample}) for the \rpfifty and \rpthirty pairs in each of our three redshift ranges.

\subsection{Isolated Sample} \label{subsec:iso_sample}
In order to test for an enhancement in star formation in close pair galaxies due to tidal interactions, we need a control sample with which to compare.  To do this, we first define a population of isolated galaxies (hereafter referred to as the ``isolated sample'') by considering only galaxies in the parent sample with no other objects within a projected separation of $300~h^{-1}$ kpc and $\Delta z / (1+z)$ $\leq 0.01$.  The selection criteria we apply to the pair sample based on the survey boundaries, conflicting slits, and $^{0}(u-r)$ color are also applied to the isolated sample.  We do not apply a strict redshift cut on the isolated sample, although we do match our pair sample to a subset of the isolated sample in both redshift and rest-frame $r$-band magnitude when constructing the control samples (\S~\ref{subsec:control_sample}).  The color-magnitude diagrams of the parent sample, pair sample, and isolated sample are shown in Figure~\ref{fig:cmd}.  There are a total of 7,890 galaxies in the isolated sample.

We note that roughly $\sim$20\% of the galaxies in both the pair and isolated samples have a potential companion projected within $50~h^{-1}$ kpc in the imaging data with a targeting magnitude (typically $r$-band or $i$-band) brighter than 22.5 that was not targeted by PRIMUS for spectroscopy.  These objects could potentially result in our pair sample containing objects in overdense regions or our isolated sample containing pair galaxies, both of which will dilute our results.  However, many of these potential companions may actually be at different redshifts than the objects in our samples, so it is likely that a much smaller fraction of our data is affected.

\subsection{Building a Fair Control Sample} \label{subsec:control_sample}
\begin{figure}
\centering
\plotone{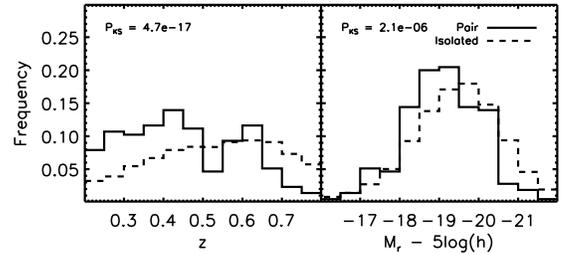}
\caption{Redshift (left) and $r$-band luminosity (right) distribution for the final \rpfifty pair sample (solid) and the full isolated galaxies sample (dashed).  The $P_{KS}$ value for each distribution is given in the top left corner of the corresponding panel.  The distributions for the two samples are clearly drawn from different populations, indicating that the full isolated galaxy sample is not a well-matched control sample for the pair galaxies.  \label{fig:pair_iso_comp}}
\end{figure}

Since PRIMUS is a flux-limited survey, close pairs detected at higher redshifts will be biased toward smaller luminosity contrasts because fainter objects will not be identified \citep{patton2000}.  At higher redshifts, isolated galaxy pairs with one lower-mass galaxy below the flux limit will be misidentified as an isolated galaxy.  To ensure that we are comparing the pair galaxies to a fair control sample, we need to create the control sample in a way such that its redshift and magnitude distributions match those of the pair galaxy sample.  The isolated sample clearly does not satisfy this, as shown in Figure~\ref{fig:pair_iso_comp}.  The numbers in the top left corner of each panel indicate the Kolmogorov-Smirnov (KS) probability, $P_{KS}$, of two distributions having been drawn from the same sample.  Furthermore, the results of simulations performed by \citet{perez2009a} suggest that 70\% of the differences between galaxies in close pairs and a control sample of isolated galaxies can be removed by constraining their redshift, stellar mass, and local environment distributions.  Much of the remaining differences between the two samples in their study can be removed by also controlling for the mass of the host dark matter halo, which we are unable to constrain with our data.  However, they claim in a subsequent study \citep{perez2009b} that they may have overestimated these effects, so their findings on the influence of host halo mass should be considered an upper limit.  We have already controlled for the local environment of our samples as described in \S~\ref{subsec:pair_sample} and \S~\ref{subsec:iso_sample}.  By also controlling for redshift and stellar mass, we eliminate the most significant biases that could influence our results.

We use a method similar to that of \citet{ellison2008} to select a subset of the isolated sample that is matched in redshift and $r$-band luminosity (as a proxy for stellar mass) to our pair sample.  For each pair galaxy, we calculate its distance from all galaxies in the isolated sample in the two-dimensional parameter space defined by the redshift and $r$-band absolute magnitude distributions of the sample.  This distance, $\Delta_{z-M_{r}}$, is defined as
\begin{equation} \label{eq:dzr}
\Delta_{z-M_{r}} = \sqrt{A^{2}(z_{pairgal}-z_{iso})^{2} + (M_{r,pairgal}-M_{r,iso})^{2}}
\end{equation}
where $z_{pairgal}$, $z_{iso}$, $M_{r,pairgal}$, and $M_{r,iso}$ are the redshifts and rest-frame $r$-band absolute magnitudes of a given pair and isolated galaxy, and $A$ is a scaling factor.  We select a fixed number, $N_{nearest}$, of isolated galaxies in order of increasing $\Delta_{z-M_{r}}$ for each pair galaxy and use those galaxies as our control sample, which results in a control sample consisting of $N_{control} = N_{nearest} N_{pairgal}$ galaxies.  By sampling with replacement, which \citet{ellison2008} did not do, our control samples can contain duplicates of the same isolated galaxy.  However, this removes any dependence of an individual pair galaxy's control sample on the rest of the pair sample.  We choose $N_{nearest}$ = 30, which is roughly the largest value such that the total number of galaxies in a control sample is no greater than twice the number of unique galaxies in that same sample across all six of the pair samples in our analysis (see below).  Based on trial and error, we set $A = 12$, which produces a consistent scaling between the values of $P_{KS}$ for the redshift and $M_{r}$ distributions with increasing $N_{nearest}$.

\begin{figure*}
\centering
\plottwo{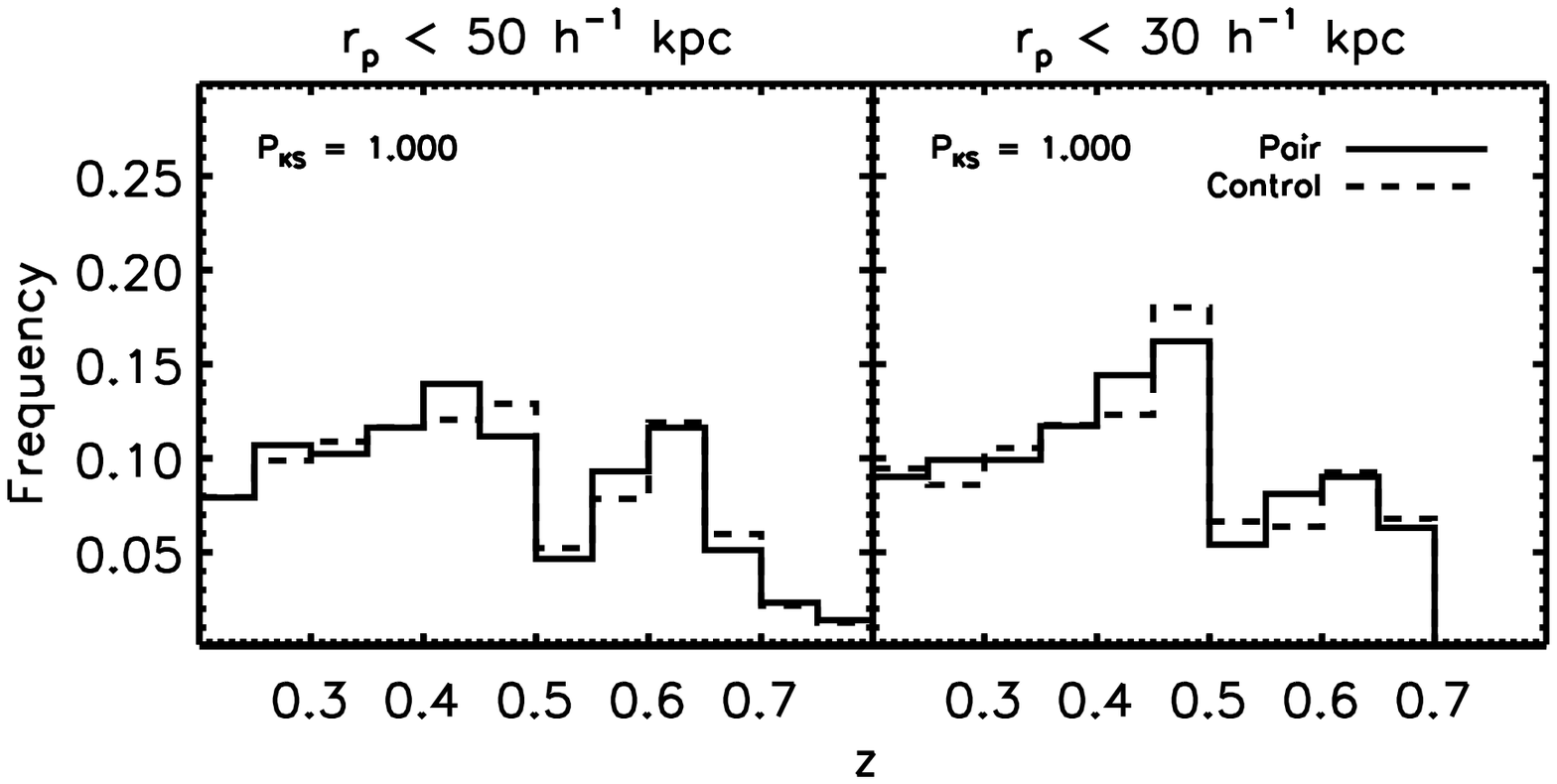}{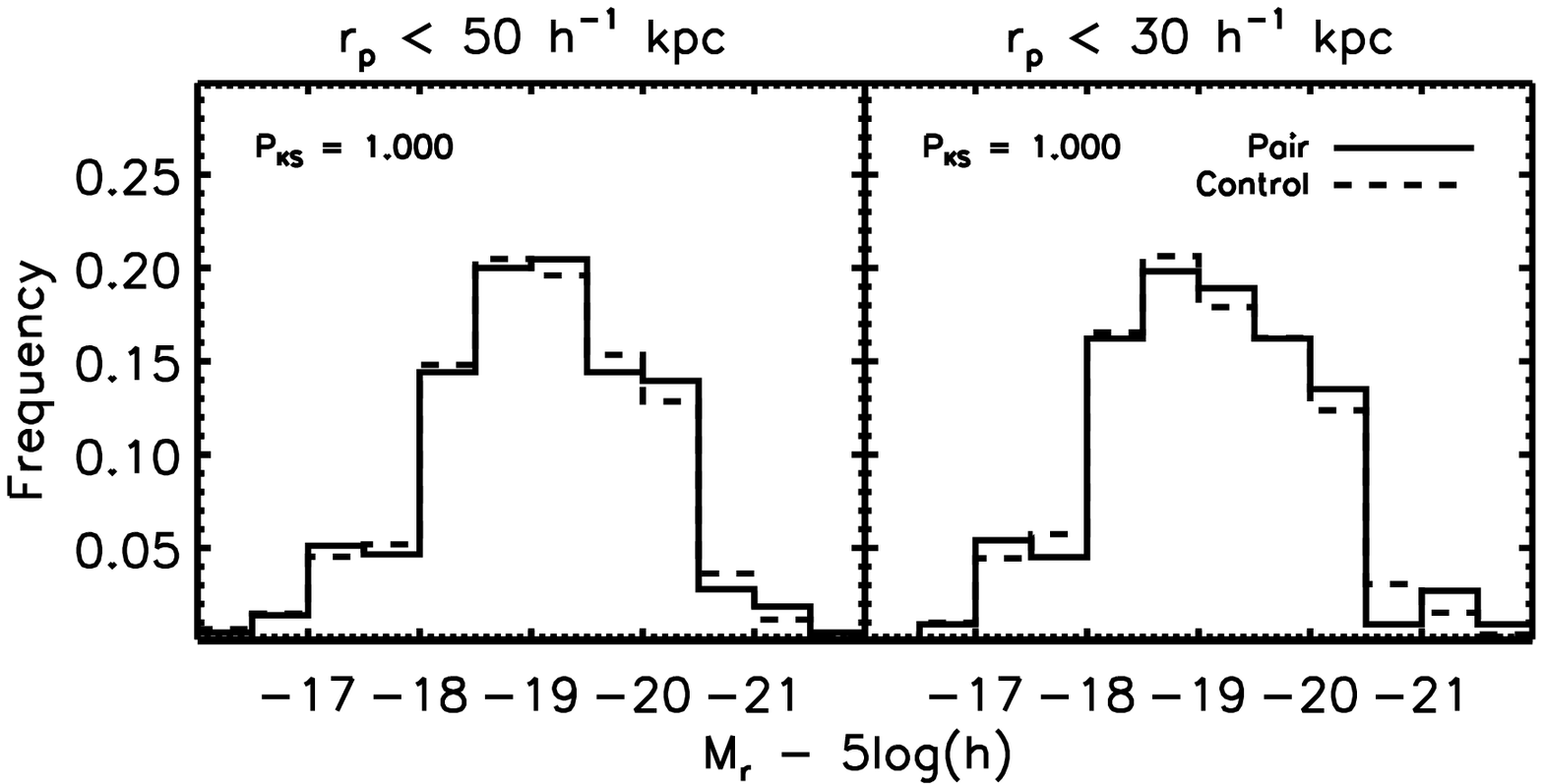}
\caption{\emph{Left:} Redshift distribution for pair galaxies (solid) and the corresponding control samples (dashed).  The two panels show the distributions for the \rpfifty pair galaxies (left) and the \rpthirty pair galaxies (right).  The KS probability of the two samples having been drawn from the same distribution is indicated in the top left corner.  In both cases, the pair and control samples are well-matched.  \emph{Right:} $r$-band luminosity distributions for pair galaxies and the corresponding control samples. \label{fig:pair_cont_comp}}
\end{figure*}

Figure~\ref{fig:pair_cont_comp} shows the redshift and $r$-band luminosity distributions for the \rpfifty and \rpthirty pair samples and their respective control samples across the full redshift range, indicating that the samples are well-matched.  We also compare the $0.25 \leq z \leq 0.5$ and $0.5 \leq z \leq 0.75$ pairs with their respective control samples and find similar results.  The number of objects in each sample for the different redshift ranges is given in Table~\ref{tab:numbers}.

\begin{table}
\begin{center}
\caption{Isolated Pair and Control Samples \label{tab:numbers}}
\begin{ruledtabular}
\begin{tabular}{c|cc|cc}
\multirow{2}{*}{Redshift} &
\multicolumn{2}{c|}{$r_{p} \leq 50~h^{-1}$ kpc} &
\multicolumn{2}{c}{$r_{p} \leq 30~h^{-1}$ kpc} \\
&
$N_{pairgal}$  &
$N_{control}$  &
$N_{pairgal}$ &
$N_{control}$
\\
\tableline
0.25 $\leq z \leq$ 0.75 &
195 &
5850 &
101 &
3030
\\
0.25 $\leq z \leq$ 0.50 &
124 &
3720 &
69 &
2070
\\
0.50 $\leq z \leq$ 0.75 &
71 &
2130 &
32 &
960
\\
\end{tabular}
\end{ruledtabular}
\end{center}
\end{table}

\section{RESULTS} \label{sec:results}
For each of our redshift ranges, we determine the distribution of $^{0}(FUV-r)$ and $^{0}(NUV-r)$ colors for the \rpfifty and \rpthirty pair samples and their respective control samples.  We calculate the shifts between the mean colors of the pair sample and its corresponding control sample,
\begin{equation} \label{eq:d_fuv_r}
\Delta\langle FUV-r \rangle = \langle ^{0}(FUV-r)_{pairgal} \rangle - \langle ^{0}(FUV-r)_{control} \rangle,
\end{equation}
and
\begin{equation} \label{eq:d_nuv_r}
\Delta\langle NUV-r \rangle = \langle ^{0}(NUV-r)_{pairgal} \rangle - \langle ^{0}(NUV-r)_{control} \rangle.
\end{equation}
The error on the mean is determined by $\sigma / \sqrt{N}$ for each sample.  To quantify the level of confidence in our result, we run a KS test to determine the likelihood of the $^{0}(UV-r)$ distributions of the pair and control samples having been drawn from the same parent distribution.

\begin{figure*}
\centering
\plottwo{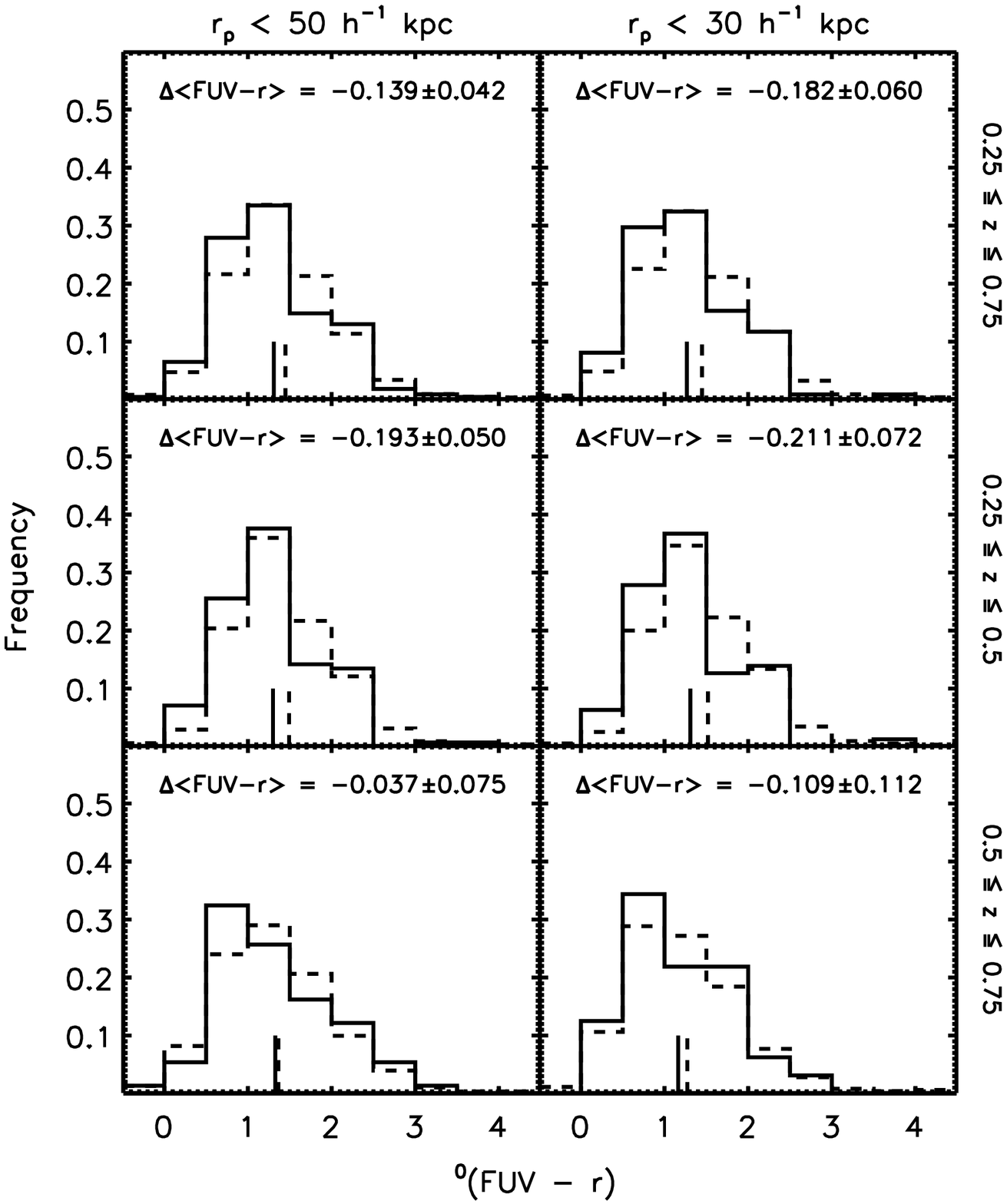}{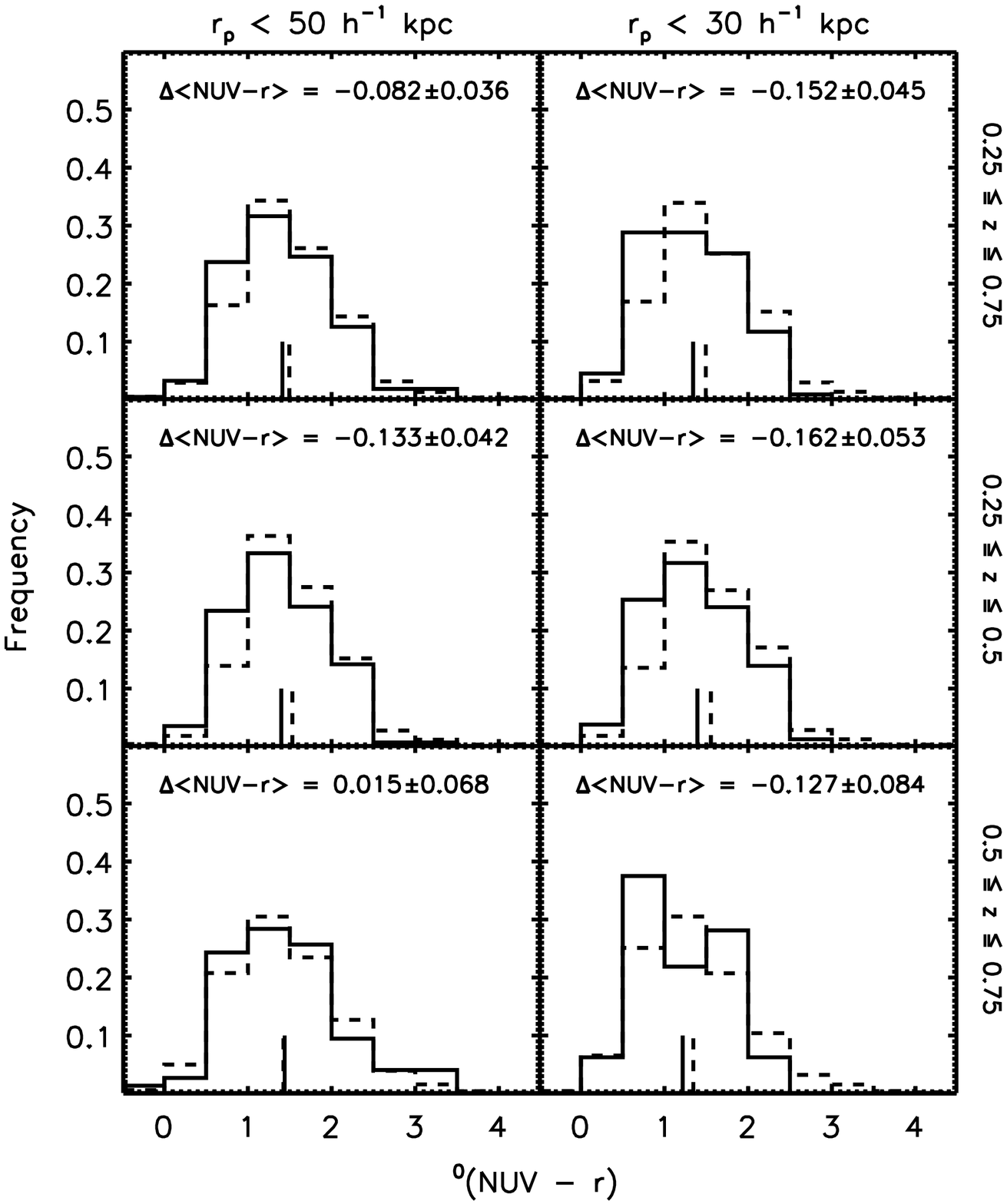}
\caption{\emph{Left:} $^{0}(FUV-r)$ distribution for pair galaxies (solid) and their corresponding control sample (dashed).  The vertical lines at the bottom of each panel indicate the mean $^{0}(FUV-r)$ color of the pair and control samples.  The two columns show the distributions for the \rpfifty pair galaxies (left) and the \rpthirty pair galaxies (right).  The three rows show the results for the full pair sample (top), the $0.25 \leq z \leq 0.5$ pair galaxies (middle), and the $0.5 \leq z \leq 0.75$ pair galaxies (bottom).  The value of $\Delta \langle FUV-r \rangle$ is given in the top left corner of each panel.  The results suggest an enhancement in the SSFR of close pair galaxies relative to the control sample at $\sim 3\sigma$ significance for the full redshift range.  \emph{Right:} $^{0}(NUV-r)$ distribution for pair galaxies and their corresponding control sample.  The results suggest an enhancement in the SSFR of close pair galaxies relative to the control sample at $\sim 2\sigma$ significance for the full redshift range. \label{fig:pair_hist}}
\end{figure*}

In the top row of Figure~\ref{fig:pair_hist}, we plot the frequency distribution of galaxies as a function of the quantities defined in Equations~\ref{eq:d_fuv_r} (left plot) and~\ref{eq:d_nuv_r} (right plot) for \rpfifty pair galaxies, \rpthirty pair galaxies, and their respective control samples of isolated galaxies.  The pair sample shows an excess of galaxies at bluer colors with a mean offset of $\Delta \langle FUV-r \rangle = -0.134 \pm 0.045$ and $\Delta \langle NUV-r \rangle = -0.075 \pm 0.038$ for the \rpfifty pair galaxies, indicating an enhancement in SSFR relative to the control sample. A stronger enhancement is seen in the \rpthirty sample with a mean offset of $\Delta \langle FUV-r \rangle = -0.193 \pm 0.065$ and $\Delta \langle NUV-r \rangle = -0.159 \pm 0.048$.  This is in agreement with past results indicating that star formation enhancement due to tidal interactions between galaxy pairs is anticorrelated with pair separation.  The values of $\Delta\langle FUV-r \rangle$ and $\Delta\langle NUV-r \rangle$ are given in Table~\ref{tab:results_ks}, along with the error bars and the corresponding values of $P_{KS}$.  The $P_{KS}$ values for the \rpfifty and \rpthirty pairs indicate that the null hypothesis (that the pair and control samples were drawn from the same $^{0}(UV-r)$ distributions) can be rejected at $\sim 95$\% significance.

\begin{table*}
\begin{center}
\caption{Results \label{tab:results_ks}}
\begin{ruledtabular}
\begin{tabular}{c|cc|cc|cc|cc}
\multirow{4}{*}{Redshift} &
\multicolumn{4}{c|}{Extinction Corrected} &
\multicolumn{4}{c}{Uncorrected} \\[1ex]
& \multicolumn{2}{c|}{$r_{p} \leq 50~h^{-1}$ kpc} &
\multicolumn{2}{c|}{$r_{p} \leq 30~h^{-1}$ kpc} &
\multicolumn{2}{c|}{$r_{p} \leq 50~h^{-1}$ kpc} &
\multicolumn{2}{c}{$r_{p} \leq 30~h^{-1}$ kpc} \\[1ex]
\cline{2-9}
&
$\Delta\langle FUV-r \rangle$ &
\multirow{2}{*}{$P_{KS}$} &
$\Delta\langle FUV-r \rangle$ &
\multirow{2}{*}{$P_{KS}$} &
$\Delta\langle FUV-r \rangle$ &
\multirow{2}{*}{$P_{KS}$} &
$\Delta\langle FUV-r \rangle$ &
\multirow{2}{*}{$P_{KS}$} \\[1ex]
&
$\Delta\langle NUV-r \rangle$  &
&
$\Delta\langle NUV-r \rangle$ &
&
$\Delta\langle NUV-r \rangle$  &
&
$\Delta\langle NUV-r \rangle$ &
\\[1ex]
\tableline
\multirow{2}{*}{0.25 $\leq z \leq$ 0.75} &
-0.134 $\pm$ 0.045 &
4.7e-02 &
-0.193 $\pm$ 0.065 &
5.0e-02 &
-0.220 $\pm$ 0.058 &
1.1e-02 &
-0.254 $\pm$ 0.087 &
3.4e-02
\\[1ex]
&
-0.075 $\pm$ 0.038 &
7.2e-02 &
-0.159 $\pm$ 0.048 &
3.8e-02 &
-0.132 $\pm$ 0.042 &
2.8e-02 &
-0.194 $\pm$ 0.058 &
2.6e-02
\\[1ex]
\tableline
\multirow{2}{*}{0.25 $\leq z \leq$ 0.50} &
-0.191 $\pm$ 0.055 &
3.8e-02 &
-0.232 $\pm$ 0.080 &
3.9e-02 &
-0.267 $\pm$ 0.071 &
1.7e-02 &
-0.334 $\pm$ 0.104 &
2.0e-02
\\[1ex]
&
-0.124 $\pm$ 0.045 &
4.2e-02 &
-0.174 $\pm$ 0.058 &
7.0e-02 &
-0.177 $\pm$ 0.051 &
1.6e-02 &
-0.236 $\pm$ 0.069 &
4.9e-03
\\[1ex]
\tableline
\multirow{2}{*}{0.50 $\leq z \leq$ 0.75} &
-0.034 $\pm$ 0.076 &
9.7e-01 &
-0.109 $\pm$ 0.112 &
7.8e-01 &
-0.137 $\pm$ 0.099 &
4.0e-02 &
-0.083 $\pm$ 0.158 &
4.7e-01
\\[1ex]
&
0.011 $\pm$ 0.068 &
1.0e+00 &
-0.127 $\pm$ 0.084 &
6.4e-01 &
-0.055 $\pm$ 0.070 &
3.9e-01 &
-0.103 $\pm$ 0.102 &
5.0e-01
\\[1ex]
\end{tabular}
\end{ruledtabular}
\end{center}
\end{table*}

\begin{figure}
\centering
\plotone{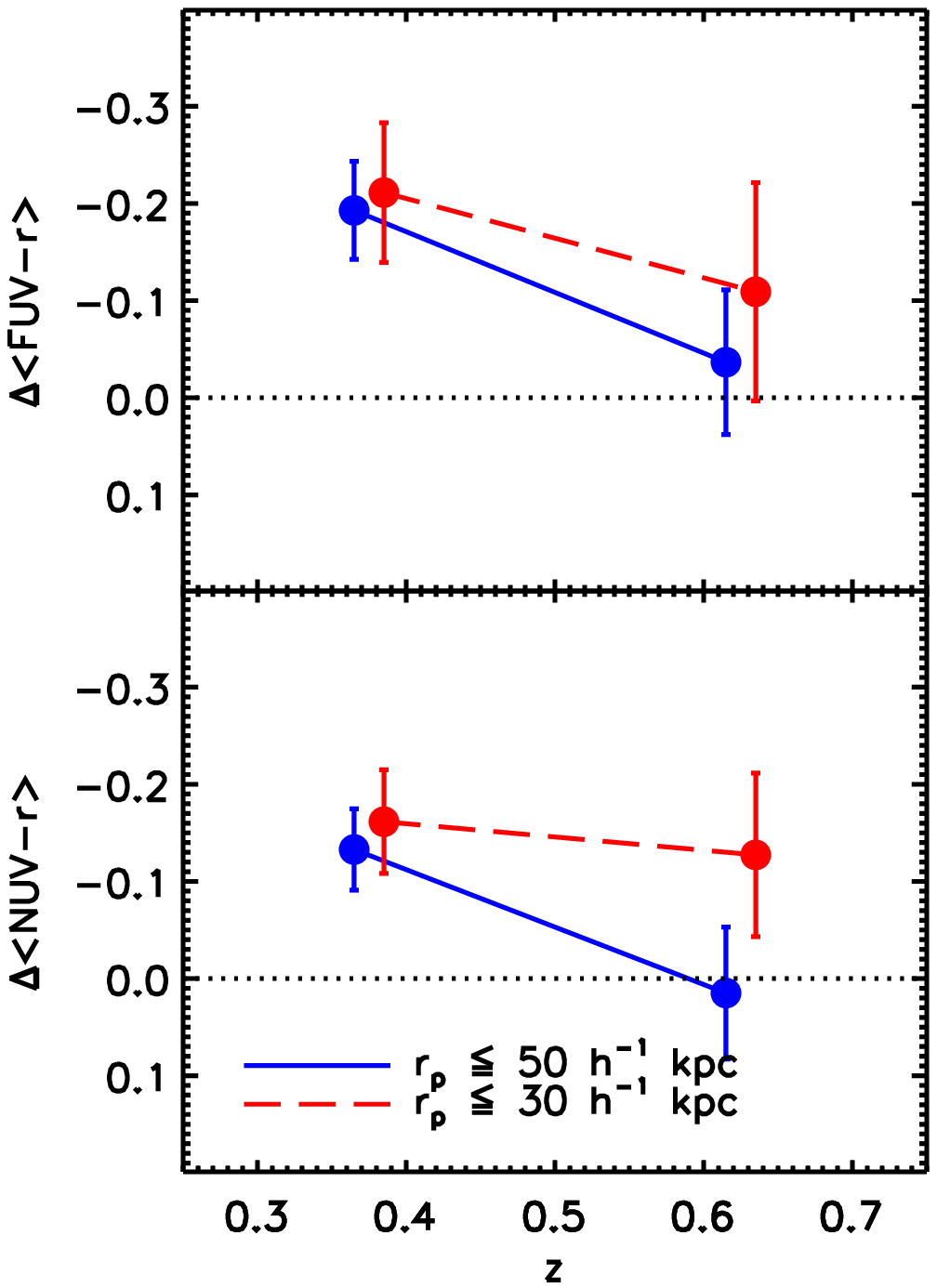}
\caption{Redshift dependence of $\Delta \langle FUV-r \rangle$ (top panel) and $\Delta \langle NUV-r \rangle$ (bottom panel) for the \rpfifty (blue solid line) and \rpthirty (red dashed line) pairs.  The points are offset from the bin centers for clarity.  The dotted line at $\Delta \langle UV-r \rangle = 0$ represents the null hypothesis of no enhanced SSFR for the pair galaxies.  We find marginal evidence for evolution of $\Delta \langle UV-r \rangle$ with redshift. \label{fig:pair_duvr}}
\end{figure}

Figure~\ref{fig:pair_hist} also shows the results for galaxies at lower ($0.25 \leq z \leq 0.5$) redshift and higher ($0.5 \leq z \leq 0.75$) redshift in the middle and bottom rows, respectively.  In the lower redshift bin, the excess of blue $^{0}(UV - r)$ colors for pair galaxies is consistent with the overall results across both $r_{p}$ samples, showing a slightly stronger enhancement for the \rpfifty sample compared to those across the full redshift range.  In the higher redshift bin, the \rpthirty pair galaxies show a weaker excess of blue colors, although this result is less significant than the lower redshift sample due to a smaller sample size (see Table~\ref{tab:numbers}).  Although a larger sample size would help to better constrain their significance, these results indicate that SSFR enhancement due to tidal interactions in small-separation pairs shows some evidence for redshift evolution between $z = 0.25$ and $z = 0.75$, although it is marginal given the error bars.  In Figure~\ref{fig:pair_duvr}, we show this main result.

\begin{figure*}
\centering
\plottwo{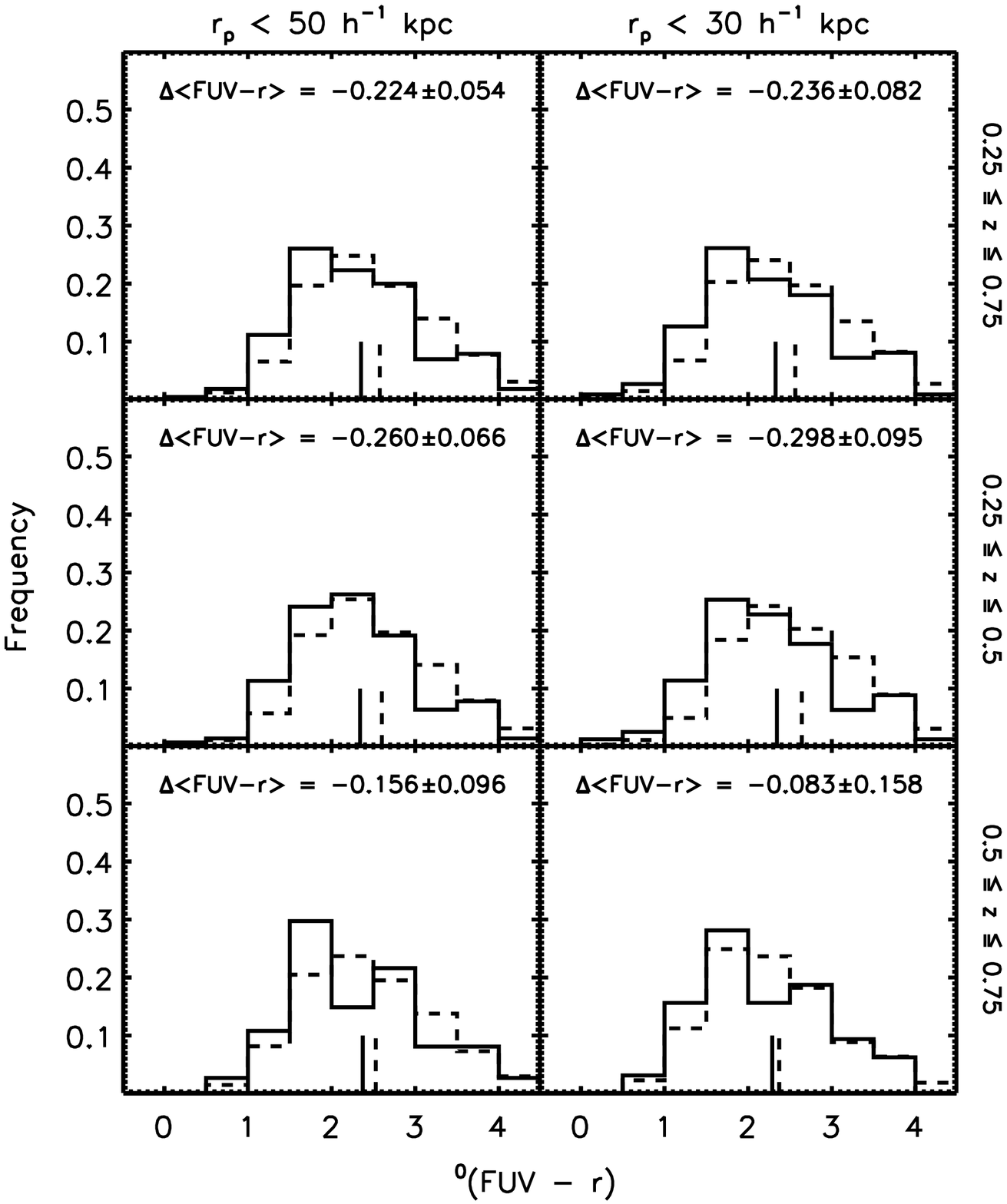}{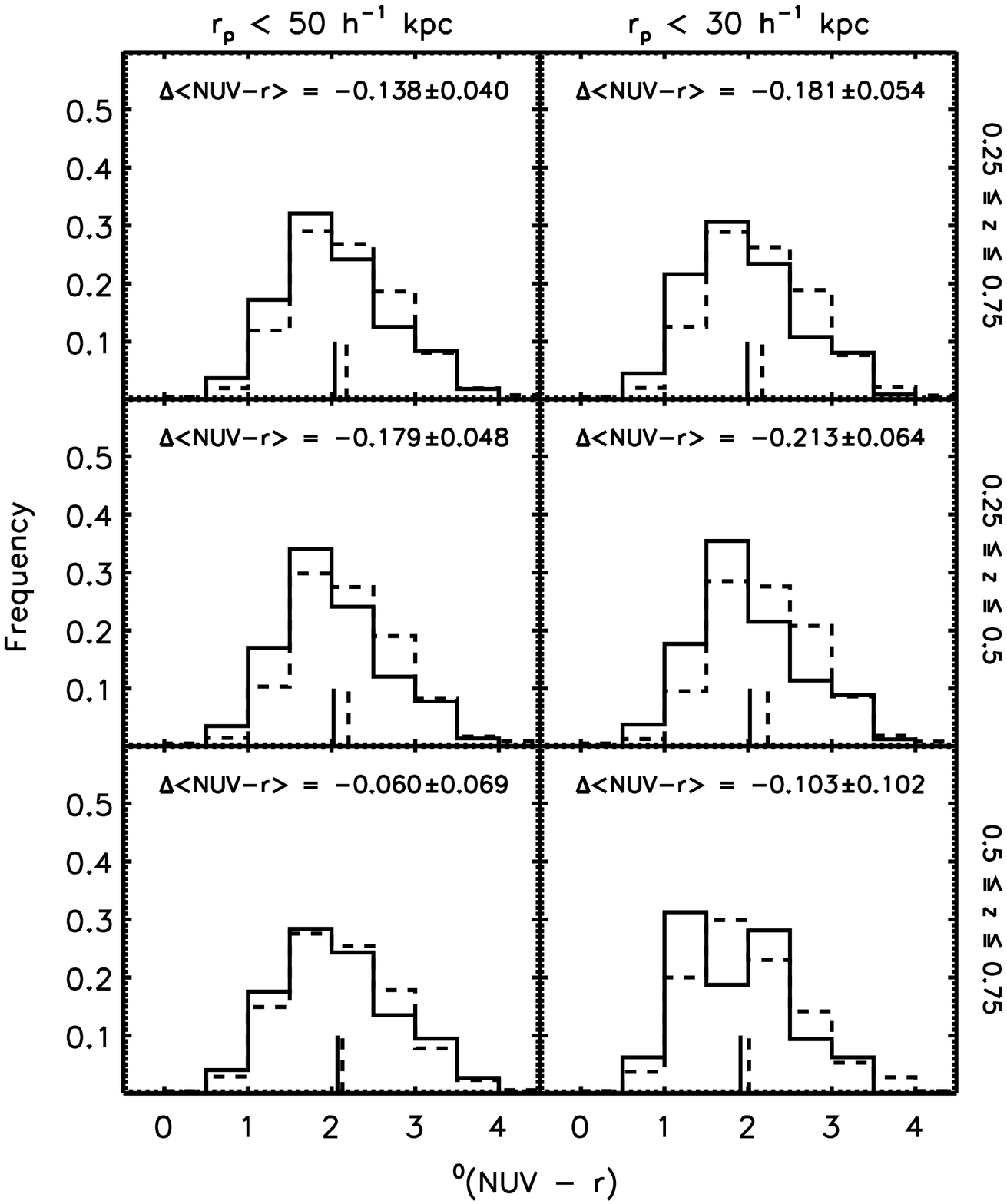}
\caption{Same as Figure~\ref{fig:pair_hist}, except without the UV extinction corrections applied.  The conclusions are similar to those in Figure~\ref{fig:pair_hist}, indicating that they are robust to the assumptions made in our extinction correction. \label{fig:pair_hist_noext}}
\end{figure*}

Given the assumptions made in performing the UV extinction correction, we also consider the uncorrected $UV-r$ colors of the pair and control samples.  Figure~\ref{fig:pair_hist_noext} shows the results obtained when the \citet{cortese2008} correction is not applied.  These results are also quantified in Table~\ref{tab:results_ks}.  We find that our general result is qualitatively unchanged when compared to the extinction-corrected results in Figure~\ref{fig:pair_hist}.  The distribution of $UV-r$ colors has shifted redward as expected, but by a similar amount for both the pair and control samples.  The $\Delta \langle FUV-r \rangle$ and $\Delta \langle NUV-r \rangle$ values have shifted slightly blueward by $\sim 1\sigma$.  The $P_{KS}$ values still indicate a result at $\gtrsim 95$\% significance.  Our conclusions are therefore robust to any uncertainties introduced by the UV extinction corrections.

\section{DISCUSSION} \label{sec:discussion}
Our study of close pair galaxies in PRIMUS reveals a stronger SSFR enhancement in the subset of \rpthirty  pair galaxies than in the subset of \rpfifty pair galaxies, which is consistent with the findings of previous studies at both low redshift \citep{kennicutt1987,barton2000,lambas2003,nikolic2004,woods2007} and intermediate redshift \citep{lin2007}.  This is likely due to the fact that at larger separations, the interacting galaxies are further away in time from perigalacticon, which is when the tidal effects between the two are greatest (tidal force approximately scales as $r^{-3}$).  Before the closest pass of the galaxies in the pair, the \rpfifty pair galaxies have had less time for tidal interactions to have affected the gas within each galaxy and triggered new star formation.  After the closest pass, the \rpfifty pair galaxies have had a longer amount of time for the increased UV flux from tidally-triggered star formation to fade due to the passive evolution of the stellar populations formed during the interaction.

To obtain a rough estimate of the fractional increase in the SSFR of pair galaxies relative to the control galaxies, we run stellar population synthesis models \citep{bruzual2003}.  To represent a galaxy before interaction with a companion, we assume a constant star formation rate of $1~M_{\odot}$ yr$^{-1}$ and run the model to $t = 6.5$ Gyr, which is approximately the time between an assumed formation redshift of $z = 3$ and the mean redshift of our sample, $z \approx 0.5$.  We also test $\tau$-models where the SFR decreases with a characteristic timescale of $\tau$ = 5 Gyr and find similar results.  The peak intensity of tidally-triggered star formation is expected to last for only a few hundred Myr \citep{dimatteo2007,dimatteo2008,woods2010}, so to simulate a tidally-triggered burst of star formation, we add a second burst of constant star formation lasting from $t = 6.5$ Gyr to $t = 6.8$ Gyr.  We then compare the $^{0}(UV-r)$ color and the SSFR  over this timescale to those of a model where the second burst was excluded (representing a control galaxy).  By running a number of trials with varying burst strengths, we determine that the mean enhancement in SSFR for the \rpfifty and \rpthirty pair galaxies is roughly $\sim 15-20$\% and $\sim 25-30$\%, respectively.  These values are broadly consistent with the results of simulations from \citet{dimatteo2008}, who find that the median integrated star formation rates of interacting galaxies are $\sim 15 - 25$\% greater than those of isolated galaxies (depending on disk inclination) over the same time period following a tidal interaction for merging systems.  Our results are also consistent with the enhancement in birthrate parameter in low-redshift galaxy pairs found by \citet{lambas2003} after accounting for the fact that they look at pairs with smaller projected separations ($r_{p} \leq 25~h^{-1}$ kpc) than we do.  We note that we have not explicitly controlled for the relative masses/luminosities of the interacting pair galaxies, so the inclusion of minor interactions in our sample may reduce the level of the average enhancement we would otherwise find \citep[e.g.][]{woods2007,ellison2008}.  The SSFR enhancement could also be biased toward lower values by interlopers in our pair sample, as projection effects can result in a significant fraction ($\sim 50$\%) of apparent pairs not being physically associated, even when a strict cut on recessional velocity difference is applied \citep{patton2008}.

Due to the likely increased gas content and frequency of tidal interactions at higher redshifts, we may have expected an enhancement of tidally-triggered star formation in close pairs to contribute to the increase in the global star formation rate at higher redshift compared to the present epoch \citep{lilly1996,madau1996,chary2001,perezgonzalez2005}.  The lack of significant redshift evolution in our results suggests that tidally-triggered star formation in close pairs is not a major contributor to the change in the global star formation rate across this redshift range.  If one assumes a strong redshift dependence of the pair fraction of $(1+z)^{4.74}$ for late-type galaxies at comparable luminosities to our sample \citep{deravel2009}, having no evolution in the enhancement of the SSFR due to tidal interactions means that roughly twice the amount of tidally triggered star formation per unit comoving volume is occurring at $0.5 \leq z \leq 0.75$ than at $0.25 \leq z \leq 0.5$.  If this enhancement is only $\sim25$\% of the SSFR of isolated galaxies (ignoring the likely lower enhancement in early-type galaxies), and pair galaxies only comprise some fraction of the entire galaxy population at a given redshift, then the total decrease in the SFR density due to tidal interactions is not a significant factor in the change of the global SFR density, which has declined by a factor of $\sim2$ over this redshift range assuming $\rho_{SFR} \propto (1+z)^{4.0}$ \citep{perezgonzalez2005}.  This implies that previous results showing that merger-induced star formation is also not a major contributor \citep{wolf2005,robaina2009,hopkins2010} can be extended to more frequent tidal interactions.  Furthermore, tidally-triggered star formation does not appear to be sensitive to changes in the gas content of galaxies as a function of redshift since we see no evidence for evolution.

Because the parent sample is drawn from a flux-limited survey, one potential source of bias is the fact that lower luminosity objects will not be detected at higher redshifts.  To test this, we run our analysis on a subset of the parent sample, including only objects with $M_{B} - 5\log{h} \leq -18.5$, which is roughly the lowest $B$-band luminosity of galaxies in the parent sample at $z = 0.75$.  We find that the SSFR enhancements in the pair galaxies shift downward by $\sim 1\sigma$, although the error bars are larger due to the smaller sample size.  There is still only marginal evidence for redshift evolution, indicating that our use of a flux-limited sample does not change our conclusions.

\section{CONCLUSIONS} \label{sec:conclusions}
Using spectroscopic data from the recently-completed Prism Multi-Object Survey (PRIMUS), in addition to existing broadband photometric data and deblended $GALEX$ data in the PRIMUS fields, we study the enhancement of SSFR in isolated close pairs of galaxies relative to a fair control sample in the redshift range $0.25 \leq z \leq 0.75$.  In comparison to the numerous investigations of tidally-triggered star formation in close pairs at low redshift ($z \sim 0.1$), few studies of a similar nature have been performed at these intermediate redshifts.  This has been primarily due to the lack of spectroscopic redshifts at the depth and volume needed to obtain a sufficiently large sample size for these studies.  PRIMUS has now allowed us to define a clean sample of isolated close galaxy pairs and to perform this analysis at intermediate redshift.

We define a sample of 195 galaxies that are in isolated close pairs with projected separation \rpfifty and $\Delta z/(1+z) \leq 0.01$, as well as a population of 7,890 isolated galaxies from which we build representative control samples that are matched in redshift and $r$-band luminosity to the pair galaxies.  We split the pair sample into two bins in redshift to further investigate the effect of redshift on the relative enhancement.  For each redshift range, we also select a subsample of \rpthirty pairs to search for a stronger signal at smaller projected separation, as is expected based on previous observational results at lower redshift.

Our results show that for the full redshift range ($0.25 \leq z \leq 0.75$), the pair galaxies have bluer $^{0}(UV-r)$ colors on average than isolated galaxies with similar redshift and $r$-band luminosity distributions by $-0.134 \pm 0.045$ magnitudes in $^{0}(FUV-r)$ and $-0.075 \pm 0.038$ magnitudes in $^{0}(NUV-r)$ for the \rpfifty pair galaxies.  For the subset of \rpthirty pair galaxies, the colors are bluer by $\Delta \langle FUV-r \rangle = -0.193 \pm 0.065$ and $\Delta \langle NUV-r \rangle -0.159 \pm 0.048$ magnitudes.  This indicates an enhancement in SSFR of roughly $\sim 15-20$\% and $\sim 25-30$\%, for the \rpfifty and \rpthirty pairs, respectively.  The stronger enhancement in the subset of \rpthirty pair galaxies is consistent with previous results at low redshifts.  A larger sample of pair galaxies would help to better constrain the significance of the enhancement.

We find marginal evidence for evolution of the SSFR enhancement in close pairs with redshift with the enhancement being slightly more significant at lower redshifts, indicating that a decrease in the level of tidally-triggered star formation in low-density environments is not a contributing effect to the overall decrease in global star formation rate density from $z \sim 1$ to the current epoch.

PRIMUS is one of the first redshift surveys with the width and depth needed to create a large enough sample of isolated pair and control galaxies for studies of tidally-triggered star formation at intermediate redshifts.  Future studies will require redshift surveys that probe larger volumes to obtain the requisite number statistics needed to better constrain the redshift evolution of tidally-triggered star formation in close pairs.

\acknowledgments
We acknowledge James Aird, Rebecca Bernstein, Adam Bolton, Douglas Finkbeiner, David Hogg, Timothy McKay, Sam Roweis, and Wiphu Rujopakarn for their contributions to the PRIMUS project.  We also thank Elizabeth Barton, Lihwai Lin, and Daniel McIntosh for insightful discussions regarding this project.  We would like to thank the CFHTLS, COSMOS, DLS, and SWIRE teams for their public data releases and access to early releases.  This paper includes data gathered with the 6.5 meter Magellan Telescopes located at Las Campanas Observatory, Chile.  We thank the support staff at LCO for their help during our observations, and we acknowledge the use of community access through NOAO observing time.  Funding for PRIMUS has been provided by NSF grants AST-0607701, 0908246, 0908442, 0908354, and NASA grant 08-ADP08-0019.  Support for this work was provided by NASA through Hubble Fellowship grant HF-01217 awarded by the Space Telescope Science Institute, which is operated by the Associated of Universities for Research in Astronomy, Inc., for NASA, under contract NAS 5-26555.  We acknowledge NASA's support for the $GALEX$ mission, developed in cooperation with the Centre National d'Etudes Spatiales of France and the Korean Ministry of Science and Technology.

\bibliography{ms}

\end{document}